\documentclass[a4paper,12pt,preprint]{elsarticle}

\usepackage{lineno,hyperref}
\modulolinenumbers[5]

\journal{SoftwareX}

\usepackage{hyperref}
\usepackage{graphicx,subcaption}
\usepackage{placeins}
\usepackage{dcolumn}
\usepackage{bm}
\usepackage{mathtools} 

\usepackage[utf8]{inputenc}
\usepackage[T1]{fontenc}
\usepackage{float}

\newcommand{\parh}[1]{\left( #1 \right) }

\renewcommand{\sin}[2][]{\mathrm{sin}^{#1}\parh{#2}}
\renewcommand{\cos}[2][]{\mathrm{cos}^{#1}\parh{#2}}

\usepackage{color}

\definecolor{LinkColor}{RGB}{48, 67, 163}

\usepackage{listings}
\newcommand{\code}[1]{\texttt{\lstinline{#1}}} 

\newcommand{\DMCPY}{\texttt{DMCPy}}

\usepackage{standalone}
\usepackage{tikz}
\usetikzlibrary{shapes.callouts,calc,arrows,shapes,positioning,shapes,arrows,fadings,decorations.pathreplacing,shadows}
\usetikzlibrary{fadings,through,angles,quotes,decorations.pathmorphing,patterns}
\tikzset{snake it/.style={decorate, decoration=snake}}
\usepackage{ifthen}

\begin{document}

\begin{frontmatter}

\title{DMCpy: A powder and single crystal neutron diffraction software for DMC}


\author[psi,danscatt]{Jakob Lass}
\author[psi,ife]{Samuel Harrison Moody}
\author[psi,ife]{\O ystein Slagtern Fjellv\aa g}

\address[psi]{PSI Center for Neutron and Muon Sciences, 5232 Villigen PSI, Switzerland}
\address[danscatt]{Nanoscience Center, Niels Bohr Institute, University of Copenhagen, 2100 Copenhagen {\O}, Denmark}
\address[ife]{Department for Hydrogen Technology, Institute for Energy Technology, NO-2027 Kjeller, Norway}
\cortext[coor]{
\textit{E-mail address}: jakob.lass@psi.ch}

\begin{abstract}

The recently upgraded DMC diffractometer at SINQ, equipped with a state-of-the-art 2D He detector, enables high-resolution neutron diffraction experiments optimized for both powder and single-crystal studies. To address the increased complexity and volume of data produced by this instrument, we developed \DMCPY{}, a Python-based software package tailored specifically for DMC data analysis. \DMCPY{} facilitates seamless data reduction and visualization, supporting conversion to reciprocal space, normalization, and masking of detector artifacts. Its modular architecture integrates tools for analyzing both powder diffraction patterns and single-crystal datasets, including advanced visualization features like 3D reciprocal space mapping and interactive scan inspection. By streamlining workflows and enhancing data interpretation, \DMCPY{} empowers researchers to unlock the full potential of the DMC instrument for probing nuclear and magnetic structures in condensed matter systems.
\end{abstract}


\end{frontmatter}


\section*{Current code version}
\label{sec:CurrentCodeVersion}

\begin{table}[H]
\begin{tabular}{|l|p{6.3cm}|p{6.3cm}|}
\hline
\textbf{Nr.} & \textbf{Code metadata description} & \textbf{DMCpy}  \\
\hline
C1 & Current code version & 1.0.1  \\
\hline
C2 & Permanent link to code/repository used for this code version & \url{https://doi.org/10.5281/zenodo.13843767}  \\
\hline
C3 & Code Ocean compute capsule & N/A \\
\hline
C4 & Legal Code License   & Mozilla Public License 2.0 (MPL-2.0)  \\
\hline
C5 & Code versioning system used & git  \\
\hline
C6 & Software code languages, tools, and services used & Python, scipy, matplotlib  \\
\hline
C7 & Compilation requirements, operating environments \& dependencies &  \\
\hline
C8 & If available Link to developer documentation/manual &  \url{https://dmcpy.readthedocs.io/}  \\
\hline
C9 & Support email for questions & \url{mjolnirpackage@gmail}.com  \\
\hline
\end{tabular}
\caption{Code metadata (mandatory)}
\label{tab:Codemetadata} 
\end{table}


\section{Introduction}

Neutron diffraction, both for powders and single crystals, is a powerful technique for probing short- and long-range nuclear and magnetic structures \cite{Lovesey}. Traditionally, powder and single-crystal diffraction instruments were designed with distinct purposes, limiting their interchangeability. However, advancements in detector technology, particularly the adoption of 2D read-out systems, have enabled the development of versatile instruments capable of efficiently handling both experimental modes. \

Notable examples of modern 2D neutron diffractometers include WOMBAT at ANSTO \cite{Avdeev2009}, WAND2 at Oak Ridge \cite{Frontzek2018}, WISH at ISIS \cite{Chapon2011}, and XtremeD at ILL \cite{Rodriguez-Velamazan2011}. Joining these, the recently upgraded DMC diffractometer now boasts a large-area $^3$He detector offering 2D read-out capabilities. Optimized for a high flux of cold neutrons, DMC’s focus provides a specialized tool for magnetic neutron diffraction, setting it apart from other constant-wavelength instruments. \

The data reduction workflow for single-crystal experiments at DMC differs significantly from traditional single-crystal instruments, particularly due to its constant-wavelength operation and large 2D detector. A typical experiment involves mapping extensive areas of reciprocal space around a single scattering plane. These measurements aim to identify magnetic Bragg scattering—which reveals magnetic ordering vectors—or to characterize diffuse magnetic scattering, often used in Reverse Monte Carlo or mean-field modeling \cite{Paddison2013,Paddison2023}. Given the dual capabilities of DMC, the data reduction process must support both standardized powder diffraction and efficient conversion of single-crystal data into Q- and HKL-space for subsequent analysis. \

To address the specific requirements of the DMC instrument, we developed the \DMCPY{} software package. Designed for flexibility and precision, DMCPy facilitates custom data reduction workflows tailored to DMC’s capabilities. It includes easy-to-use and robust tools for processing both powder diffraction and single-crystal data, ensuring seamless analysis across a wide range of experiments. \

\subsection{The DMC diffractometer}

The DMC diffractometer, located at SINQ, is designed for high-flux at low-Q neutron diffraction experiments. It specializes in leveraging cold neutrons to investigate both nuclear and magnetic structures in condensed matter systems. The instrument’s flexibility allows it to accommodate a broad range of experimental setups, from standard powder diffraction to advanced single-crystal studies. \

A key feature of the DMC instrument is its large-area $^3$He detector. This advanced 2D detector enables the collection of scattering data over a wide angular range, significantly enhancing the resolution and efficiency of experiments. The detector’s cylindrical geometry spans a vertical coverage of 20 cm and a horizontal arc of 130$^\circ$, allowing comprehensive sampling of reciprocal space. With an 80 cm detector-sample distance, the setup provides an out-of-plane angular resolution of ±7.13$^\circ$, ideal for capturing complex scattering phenomena.

\begin{figure}[ht!]
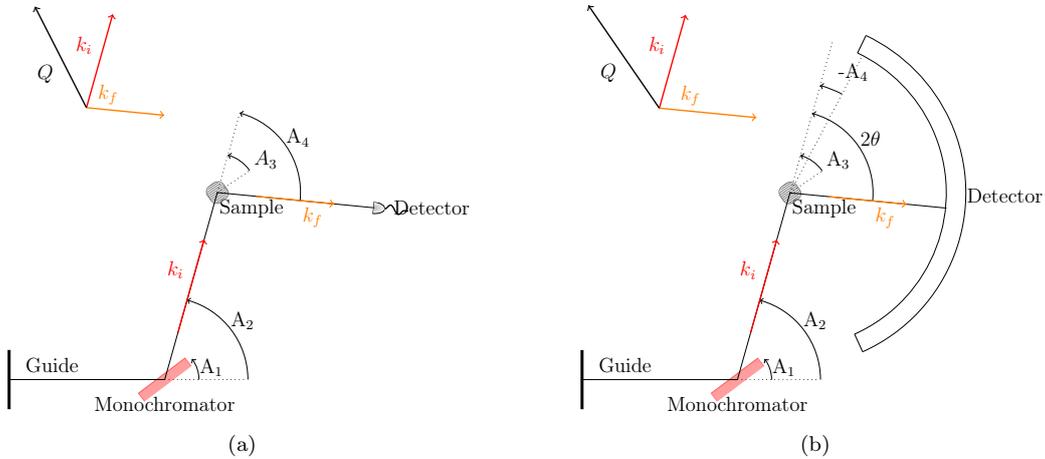

    \centering
    \begin{subfigure}[b]{0.45\linewidth}
    \includestandalone[width=\linewidth]{figures/Instrument/DoubleAxisInstrument}
    \caption{} 
        \label{fig:DoubleAxisInstrument}
    \end{subfigure}\hfill
    \begin{subfigure}[b]{0.45\linewidth}
    \includestandalone[width=\linewidth]{figures/Instrument/DoubleAxisInstrument2}
    \caption{} 
        \label{fig:DMCInstrument}
    \end{subfigure}
    \caption{(\textbf{a}): Sketch of standard double axis diffractometer. (\textbf{b}): Sketch of DMC with a large detector coverage highlighting the angle conventions.}
    \label{fig:DMCOverview}
\end{figure}

The DMC diffractometer’s design ensures seamless integration between the detector’s angular coverage and the instrument’s motion capabilities (See Fig.~\ref{fig:DMCOverview}. The sample table, capable of precise rotation along the vertical axis (A3), facilitates systematic exploration of reciprocal space. Coupled with the detector’s wide angular coverage (A4 or 2$\theta$), this arrangement allows researchers to efficiently map scattering patterns in three dimensions. \

\subsection{Software introduction}

\DMCPY{} is a Python library specifically designed to address the unique requirements of the DMC diffractometer. Distributed via Python's package interface (PyPI), the software is compatible with Python versions 3.6 and later. It relies on widely used libraries such as Matplotlib \cite{Matplotlib}, Pandas \cite{PandasSoftware}, Numpy \cite{numpy} and H5py \cite{h5py}, providing a robust foundation for data analysis. \

The software is organized into six core modules, each responsible for specific aspects of data handling, visualization, and analysis. At the heart of \DMCPY{} is the \code{DataFile} object, which performs critical tasks like converting experimental data from real-space detector coordinates to reciprocal-space representations, normalizing data, and masking instrumental artifacts. The \code{Sample} object complements this functionality by converting data into sample-specific reciprocal lattice units based on input from the SICS neutron scattering library \cite{Greuter2002}. \

To efficiently manage complex datasets, the \code{DataSet} object acts as an advanced list structure, enabling simultaneous handling of multiple scans. This object is integrated with powerful visualization tools, including \code{Viewer3D} for navigating single-crystal data in three-dimensional reciprocal space and \code{InteractiveViewer} for inspecting raw detector data scan-by-scan. The software also introduces \code{RLUAxes}, a plotting tool that overlays data with reciprocal lattice grids to ensure consistent and distortion-free visualization across different space groups and unit cells. \

Lastly, the two scripts: \code{_tools} and \code{FileStructure}, have been written to collect commonly-used functions and provide a flexible data-file loading framework respectively. \DMCPY{} is distributed with the normalization data which takes care of the cross-normalization between different parts of the detector, see sec.~\ref{sec:normalization}.

\section{Functionality and Scope of \DMCPY}

\DMCPY{} is designed in a similar way to the MJOLNIR \cite{MJOLNIR} utilised at the CAMEA instrument \cite{Lass2023CAMEA}, SINQ, PSI, Villigen and provides tools for four purposes: 1) provide a quick overview of data as initially acquired facilitating alignment and for the estimation of needed counting time. 2) conversion of data into the reciprocal lattice coordinate system and the sample coordinate system. 3) Normalisation and visualisation of the data in 1, 2 and 3 dimensions, and 4) tools to aid the analysis of single-crystal data. Below we describe the scope of \DMCPY{} in these aspects.

\subsection{Raw data overview - Interactive Viewer}
\begin{figure}[ht!]
    \centering
    \includegraphics[width=0.8\linewidth]{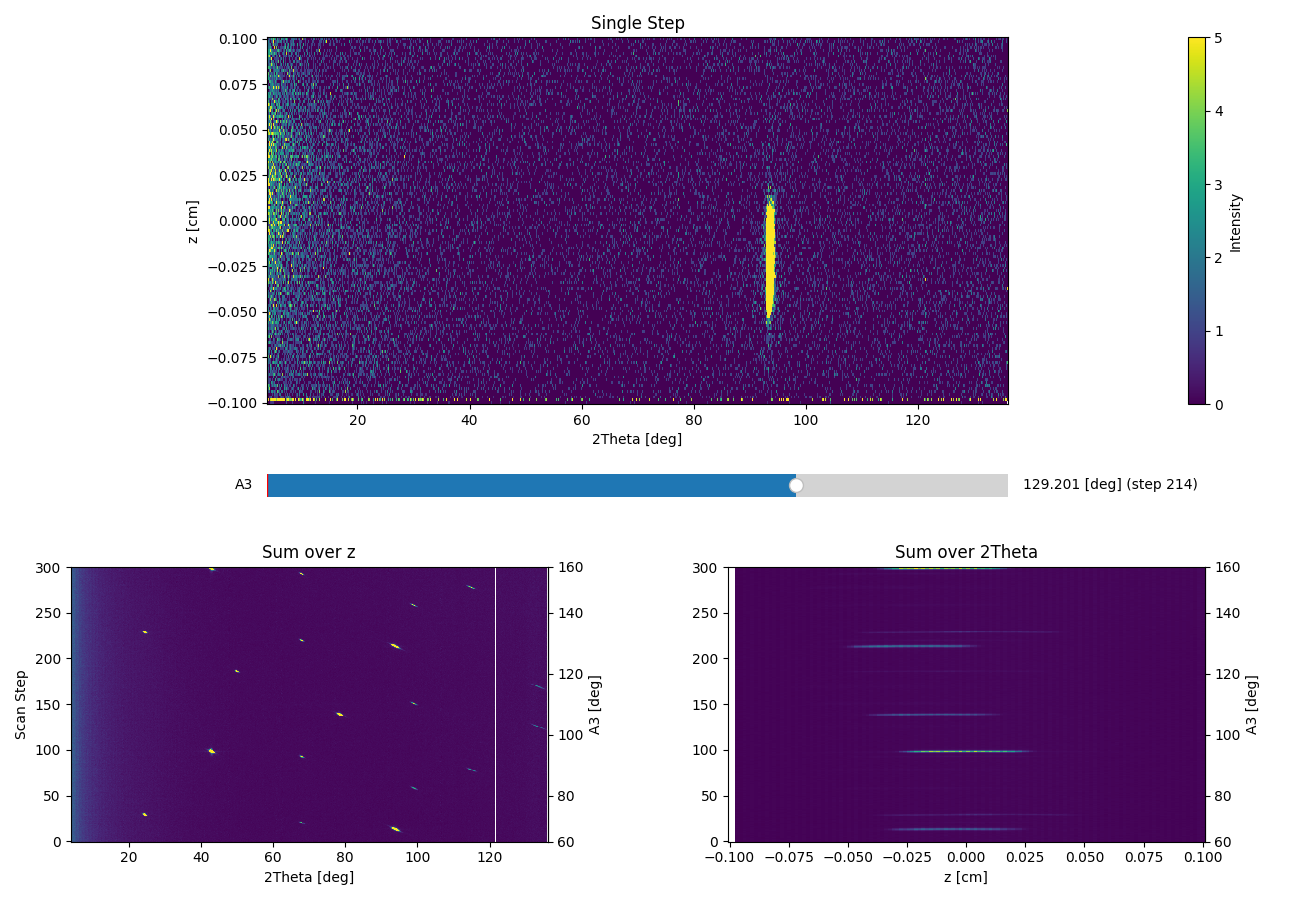}
    \caption{Interface of the Interactive Viewer. Top: The figure shows the intensity distribution at the detector for a specific $A_3$. The $A_3$ step can be chosen with the scroll bar. Bottom left: Intensity as a function of $A_3$, where the detector is summed along the z-direction (out-of-plane). Bottom right: Intensity as a function of $A_3$, where the detector is summed along two theta.}
    \label{fig:IA}
\end{figure}

While acquiring data, an overview of the collected data is needed. For the case of powder samples, this overview is created by integrating and exporting the resulting one-dimensional data. 
For single crystal samples, one needs first to identify a rough alignment. When inserting the sample in the sample environment, the exact sample rotation is unknown and it is uncertain whether the wanted scattering plan coincides with the horizontal plane of the detector. Thus, a quick sample rotation scan is performed with a fine step size to ensure that all peaks are found. To showcase such data the interactive viewer was created, Figure \ref{fig:IA}. It takes a single data file and displays three figures: (I) A plot of the intensities on the detector for a given sample rotation, $A_3$, with a slider to choose different $A_3$'s, (II) A plot of the intensity summed along z, i.e. vertically on the detector, as a function of $A_3$, and (III) A plot of the intensity summed over scattering angle, $2\theta$, as a function of $A_3$. The functionalities of the Interactive Viewer allow inspection of the crystal alignment, determination of the $A_3$ for specific Bragg peaks, etc.

\subsection{Data Conversion}
By utilising the coordinate system introduced by Lumsden \cite{Lumsden2005} the transformation from pixel positions to scattering vector transfer is performed. The sample position is chosen to coincide with the origin, $(0,0,0)$ and the incoming beam along $\vec{k}_i = (0,k,0)$, where $k$ is the length of the wave frequency of the neutron. To ensure a right-handed Cartesian coordinate system $x$ and $y$ are chosen to span the horizontal scattering plane and $z$ to point vertically upwards. The  wavelenght of the incoming neutrons selected by the monochromator is calculated through the Bragg scattering equation and the scattering angles of the monochromator. Similarly, the final neutron propagation vector, $\vec{k}_f$, is defined within the same coordinate system and points towards the individual pixels on the detector. The pixel positions are assumed to be known from the instrument geometry. This allows the direct calculation of the scattering vector $\vec{Q}$
\begin{equation}
    \vec{Q} = \vec{k}_i-\vec{k}_f.
\end{equation}
In some experiments, the centre of scattering of the sample is displaced vertically. 
For such samples, two modifications to the above are needed; first, the incoming beam no longer follow the $y$-axis but rather is angled upwards by $tan(\nu) = d_z/d_{ms}$, where $d_z$ is the vertical displacement and $d_{ms}$ is the monochromator to sample distance. The incoming scattering vector thus becomes $\vec{k}_i = k_i(0,\cos{\nu},\sin{\nu})$. Second, the pixel position on the detector have to be moved oppositely, i.e. relative to the sample, by $d_z$. 

%
%
\subsection{Projection into the Reciprocal Lattice System}
To convert between the instrument coordinate system and that of the sample, a conversion matrix, the UB matrix, is needed. It takes $\vec{Q} = (Q_x,Q_y,Q_z)$ into $\vec{HKL} = (H,K,L)$ through a combination of two transformations; a unitary transformation, $U$, defining the sample orientation relative to the instrument, and a unit cell specific $B$ matrix. 
Here, it is to be noticed, that the sample rotation, denoted $A_3$, corresponds to a simple rotation within the $Q_x$-$Q_y$ plan, i.e. orthogonal to $(0,0,Q_z)$. That is, when a sample rotation scan is performed, the scattering vectors in the sample space for all pixels can be found by rotating an initial coverage by the sample rotation around the vertical axis. In \DMCPY{}, this is leveraged through the implementation of a so-called ''lazy calcualtion'' of the scattering positions, i.e. a just-in-time (jit) calculation reducing the RAM usage significantly. 

To determine the correct UB matrix for a given crystal orientation two indexed peaks are needed most often found by hand. Alternatively, the algorithm explained in App.~\ref{sec:PeakSearch} can be used to automatically suggest peaks.

\subsection{Vanadium normalization}\label{sec:normalization}
\begin{figure}
    \centering
    \includegraphics[width=0.95\linewidth]{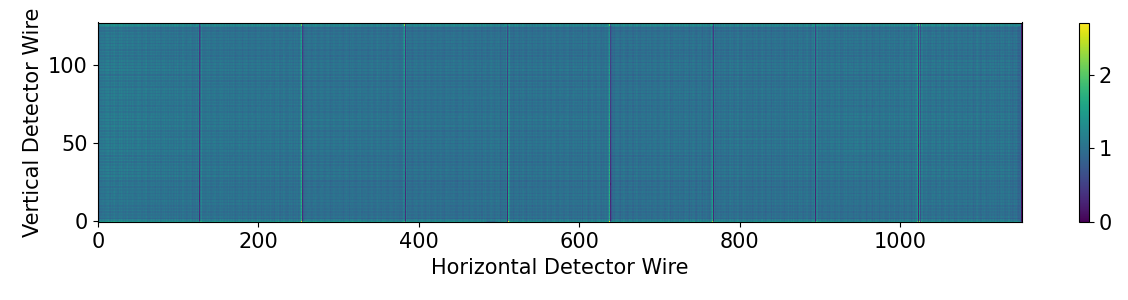}
    \caption{Vanadium normalization table for DMC data 2023.}
    \label{fig:normalization}
\end{figure}

The DMC detector consists of 9 identical modules, each with 128 x 128 wires spanning a 2D area of 20 x 20 cm$^2$ corresponding to a total coverage of 180 x 20 cm$^2$. These 9 modules are connected to pre-amplifies and are run individually through the data acquisition system. This results in discrepancies in the sensitivity both within individual modules and, in particular, between different modules. To alleviate this, the standard method of a normalization sample is utilized to perform a gain correction. For DMC, a Vanadium sample 
is measured for a around two hours. The pixel by pixel values are divided by the median value producing a normalization table with an average sensitivity of 1. This table is then saved and uploaded to the \DMCPY{} repository making it available in all future versions. Masking of certain areas, wires, or even modules can be done by substituting the corresponding values in the normalization table by NaN, i.e. Not-a-Number, values. This propagates through the program and masks out these detector pixels in such a way that they are ignored. An example of the normalization table from 2023, highlighting the different sensitivity of the individual tubes, is shown in Figure~\ref{fig:normalization}. The repetitive motive of 128 wires per module can be seen by the large vertical difference while a much smaller variance can be seen between individual wires.

\subsection{Data analysis}
To facilitate data analysis, a series of integration and export functions have been created, dealing with both powder and single crystal data. These include binning of data in 3D, 2D colour maps as well as 1D cuts, and for powder measurements constant $|q|$ integration or standard vertical integration. In addition, masking is provided allowing the removal of instrument or sample artefacts, or the removal of data with too coarse resolution, e.g. out of plane. All of these features are explained further below.

All of the binning and integration methods work on the converted, detector pixel data. These can have a considerable size in memory when large sample rotation scans have been performed and multiple detector positions used. For a single sample rotation scan, the number of data points measured are $n_s \times 128 \times 1152 $, where $n_s$ is the number of steps in the scan, 128 and 1152 signify the number of vertical and horizontal pixels. For a standard scan covering 60 degrees in steps of 0.1 degree with two detector positions, the total number of pixels exceeds 88 million. Performing even simple operations on such a block of data using a standard laptop is unfeasible. Thus, all of the data analysis methods in \DMCPY{} are setup to allow for batched computation. That is, only part of a data file is treated at once and the algorithms step through parts of files sequentially. This imposes restrictions on the possibility to manipulate data,  e.g. the subtraction of two data sets requires the creation of a third into which the difference is stored. In addition, when a UB matrix is found, it needs to be saved into the working data file as well. 

\section{Powder Data Functionality} 
This section will address the capabilities of \DMCPY{} for data reduction of powder data.
\subsection{Data integration}
\begin{figure}[ht!]
    \centering
    \includegraphics[width=0.95\linewidth]{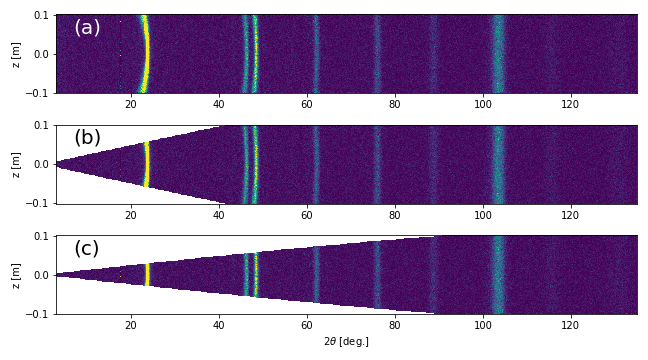}
    \caption{Gain-corrected intensity of one detector image with (\textbf{a}) no mask, (\textbf{b}) a 10 deg. mask, and (\textbf{c}) a 5 deg. mask for an MnS sample measured at 70 K with $\lambda$ = 2.46 \AA. }
    \label{fig:plotDetector}
\end{figure}

The DMC detector collects 2D neutron counts which can be visualized as intensities on the detector through the \code{plotDetector} function. An example of a standard diffraction pattern using MnS is shown in Figure \ref{fig:plotDetector}. For integrating the data into 1D diffraction patterns, \DMCPY{} calculates the length of Q for each detector pixel before projecting the data into bins. The bins are by default calculated with a step-size of 0.125\textdegree , however, other step sizes can be defined and custom bins can also be defined by the user. 

\begin{figure}[ht!]
    \centering
    \includegraphics[width=1.0\linewidth]{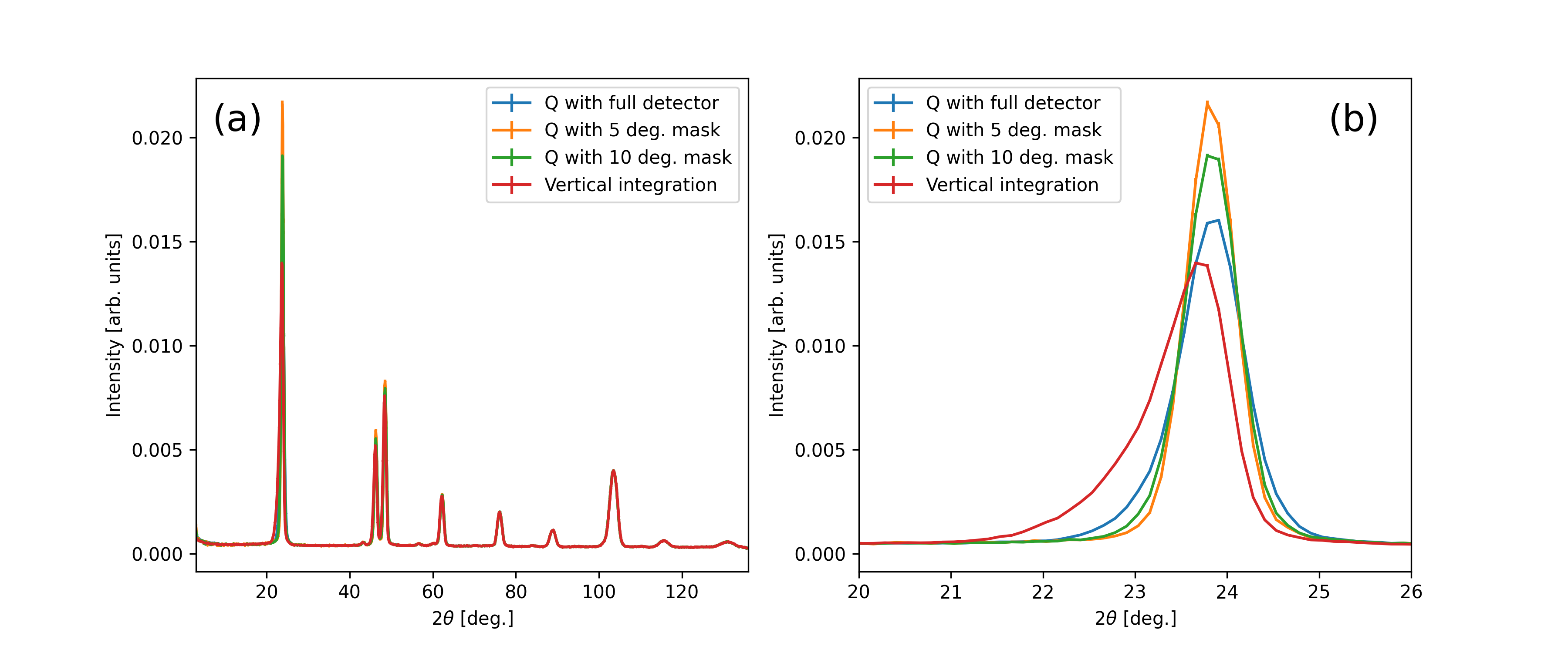}
    \caption{Integrated diffraction patterns of an MnS sample measured with six detector positions at 70 K and $\lambda$ = 2.46 \AA, as shown in Figure \ref{fig:plotDetector}. (\textbf{a}) Full diffraction pattern, and (\textbf{b}) zoomed area which illustrate the difference in peak shape and resolution at low angles depending on the integration procedure and mask selection.  }
    \label{fig:plotTwoTheta}
\end{figure}

Instead of the proper calculation of the length of the scattering vector, due to backwards capability, it is further possible to perform a vertical integration of the data where the out-of-plan scattering vector component is disregarded. Historically, such an integration is combined with a constant out-of-plane scattering angle mask, as shown in Figure \ref{fig:plotDetector}. A comparison of integration in Q-space and vertical integration is shown in Figure \ref{fig:plotTwoTheta}. Visual inspection of the data clearly show that |Q|-integration produces more symmetrical peaks as compared to vertical integration. For the latter, the centre of mass of the reflection is shifted to a lower angle, when below 90 \textdegree and higher angle when above, due to the curvature of the powder rings, exacerbated at both low and high scattering angles. It is also evident that the powder rings are broader out-of-plane at low angles than in-plane. The broadening is consistent with the focussing monochromator at DMC. To increase the resolution at a low angle, it is possible to use an angular mask to remove the broader part of the powder ring. Figure \ref{fig:plotDetector} and Figure \ref{fig:plotTwoTheta} also show how a 5 and 10 deg. mask remove data and increase the resolution. However, we note that masking out parts of the detector to increase the resolution will correspondingly reduce the statistics. 

\subsection{Powder Data formats}
\DMCPY{} offers efficient integration of data and exports the data as PSI-powder format or using the xye-format, both directly importable in standard powder software, e.g. FullProf \cite{Rodriguez-Carvajal1993}, GSAS-II \cite{Toby2013} Jana2020 \cite{Petricek2023}. This is done using either the \code{export} or \code{sortExport} functions, which in turn runs \code{export_PSI_format} and \code{export_PSI_format}. \code{sortExport} merges files with the same title and is very efficient for combining data files in an experiment. In \code{export}, data files that should be merged need to be defined by the user. Combining data files is important to average data with different detector positions to remove detector artifacts not fully removed by normalization or to improve statistics.

\subsection{Validation}
The powder features of \DMCPY{} have been thoroughly validated by studies published in scientific journals, where the data is analyzed by the Rietveld method \cite{Thogersen2023,Sannes2023,Huangfu2024,Gauthier2024,Bertin2024,Deptuch2024}. In addition, Rietveld refinements of NAC ($\mathrm{Na_2 Ca_3 Al_2 F_{14}}$) for calibration of the instrument illustrate successful integration of DMC data.

\FloatBarrier
\section{Single Crystal Functionality} 
This section will discuss the software functionalities designed to aid single-crystal experiments and data reduction. It consists of several functions with complementary capabilities, with their properties described below. A common feature of these functions is that the binning or resolution in Q- and HKL-space is always given in Q, i.e. in the sample independent laboratory frame of reference, i.e. in units of 1/\AA. This allows an unambiguous binning in all directions for non-cubic systems as well as across samples with different unit cell sizes.  

\subsection{Box integration of peaks}
\begin{figure}[ht!]
    \centering
    \begin{subfigure}[b]{0.95\linewidth}
        \centering
        \includegraphics[width=\linewidth]{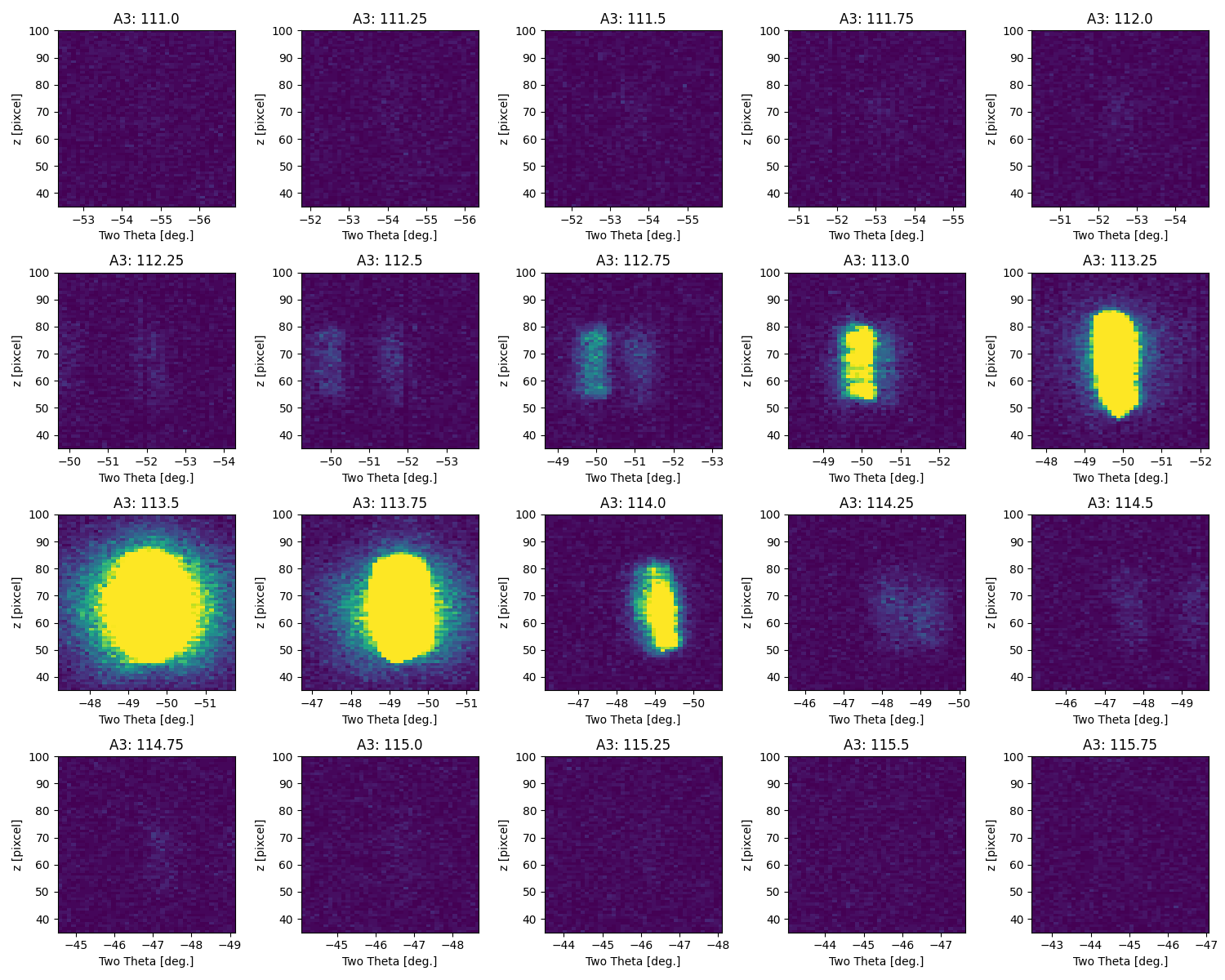}
        \caption{} 
        \label{fig:BoxIntegrationOverview} 
    \end{subfigure}
    \hfill
    \begin{subfigure}[b]{0.65\linewidth}
        \centering
        \includegraphics[width=\linewidth]{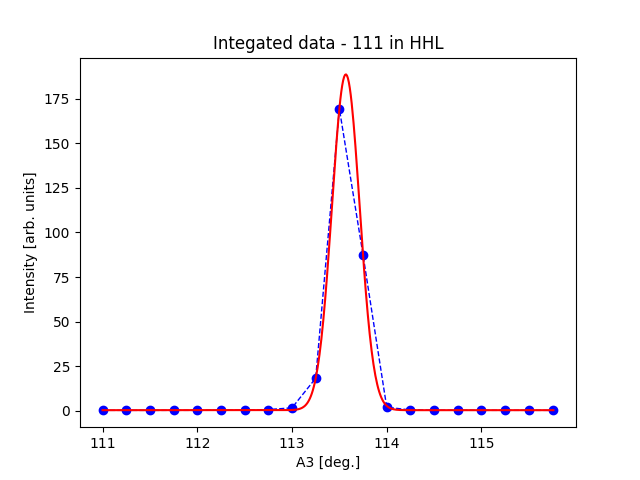}
        \caption{} 
        \label{fig:BoxIntegrationRoi} 
    \end{subfigure}
    \caption{a: Box integration performed on the (1,1,1) Bragg peak of a single crystal showing the region of interest as function of sample rotation angle. b: Summed roi as a function of $A_3$. }
    \label{fig:BoxIntegration}
\end{figure}

The standard way of performing an integration of single crystal Bragg peaks is through a so-called box integration where intensity is summed over an area on the detector as a function of sample rotation. \DMCPY{} offers an implementation of this method through the \code{boxIntegration} function. It takes a dictionary of peaks and integration parameters as input which specifies the integration widths, peak centre in scattering angle, the height on the detector, and the range around the central sample rotation value to consider. The $A_3$ and $A_4$ values of a peak can be found using the \code{calcualteHKLToA3A4Z} function or by visual inspection using the Interactive Viewer. The regions of interest (roi's) are integrated and intensities are plotted as a function of sample rotation.  an example for the (1,1,1) Bragg peak of a single crystal is shown in Figure~\ref{fig:BoxIntegration}. The roi's show that we have chosen a sensible integration range on the detector and in $A_3$. The intensity as a function of $A_3$ is also shown.

\subsection{Viewer3D}\label{sec:Viewer3D}
The quickest way to gain an overview of the collected data within the acquired 3D space is through an equi-sized binning and a method to step through sequential 2D colour maps. The \code{Viewer3D} functionality provides exactly this, either using the instrument coordinates $(Q_x,Q_y,Q_z)$ or, if a UB has been generated, the lattice specific $(H,K,L)$ projection vectors. In non-orthogonal systems or when a non-standard scattering plan has been measured, having the main axis along the exact RLU vectors is sub-optimal and an option exists to provide the projection vectors manually. If non-orthogonal vectors are provided, the viewer will still generate views along these but  binning will be performed on a orthogonal basis with the first vector along the first projection, the second vector along the cross product of projection vector 1 and 3, and the last vector similarly along the normal of the span of vector 1 and 2.

\begin{figure}[ht!]
    \centering
    \includegraphics[width=0.95\linewidth]{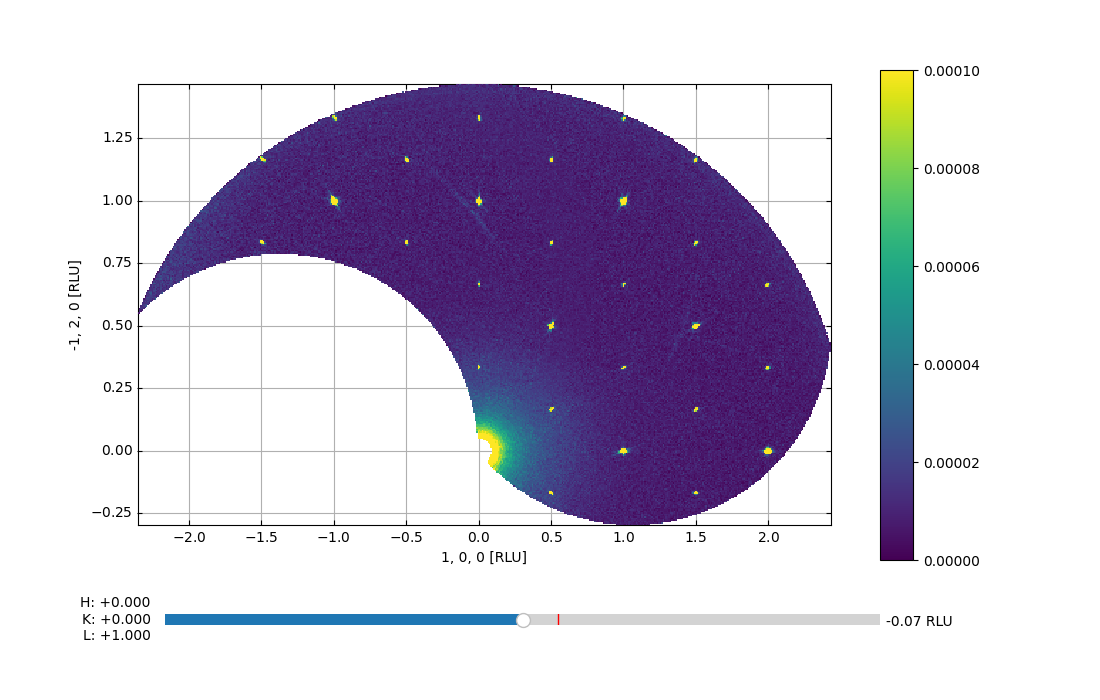}
    \caption{ The \code{Viewer3D} displays the intensity of one scattering plane in Q- or HKL-space. The scroll bar allows choosing different levels orthogonal to the displayed scattering plane. }
    \label{fig:3DViewer}
\end{figure}

Having specified the projection axes, the edge sizes of the rectangular cuboids or voxels into which the data is binned are to be provided. The binning is performed on either the un-rotated, in the case of instrument coordinate system, or the rotated data. The actual binning is performed by a modified version of the numpy \code{histogramdd} function \cite{numpy}, where multiple matrices, e.g. intensity and monitor values, are simultaneously binned using the same bin indices found from the data point positions. However, before one can bin the data the extremal axis values have to be known. This poses a problem due to the iterative data treatment which is necessary due to the data size. Naively one finding these values by first going through all of the for then in a second loop to perform the actual binning. To avoid this, the extremal values are found for on a batch by batch basis and the bin grid is updated accordingly. That is, it is insured that when two batches are combined, their grids line up perfectly. Not only does this remove the need for an extra data pass, it also minimise the RAM usage until it is actually needed, at the small cost of recalculating the grid.

The default view presented to the user is to show data with a constant value of the third axis, allowing for a change of this value by using arrow keys or mouse wheel scrolling, see Figure \ref{fig:3DViewer}. Two other views are offered, i.e. with constant projection vector 1 and 2, and one can cycle through these interactively. 

\subsection{Performing 2D and 1D cuts}
For quantitative analysis, cuts of the data into both 2D and 1D are needed. 
\DMCPY{} performs these in either the instrument or HKL space. However, as described earlier, all widths and lengths are defined in $\text{\r{A}}^{-1}$ for consistency between functions, cut directions, and between experiments. 
\subsubsection{2D Cutting}
\begin{figure}[ht!]
    \centering
    \includegraphics[width=0.95\linewidth]{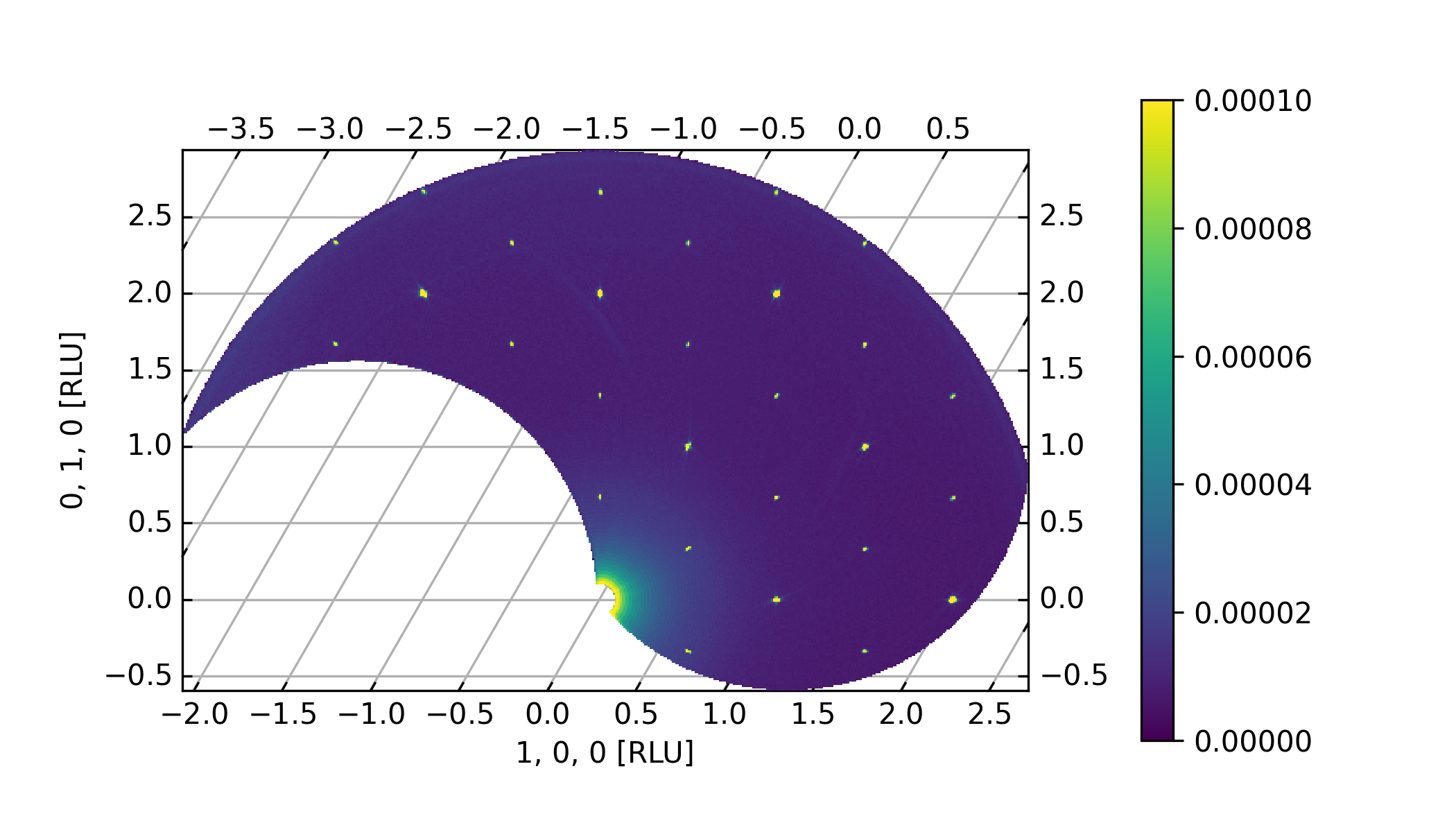}
    \includegraphics[width=0.95\linewidth]{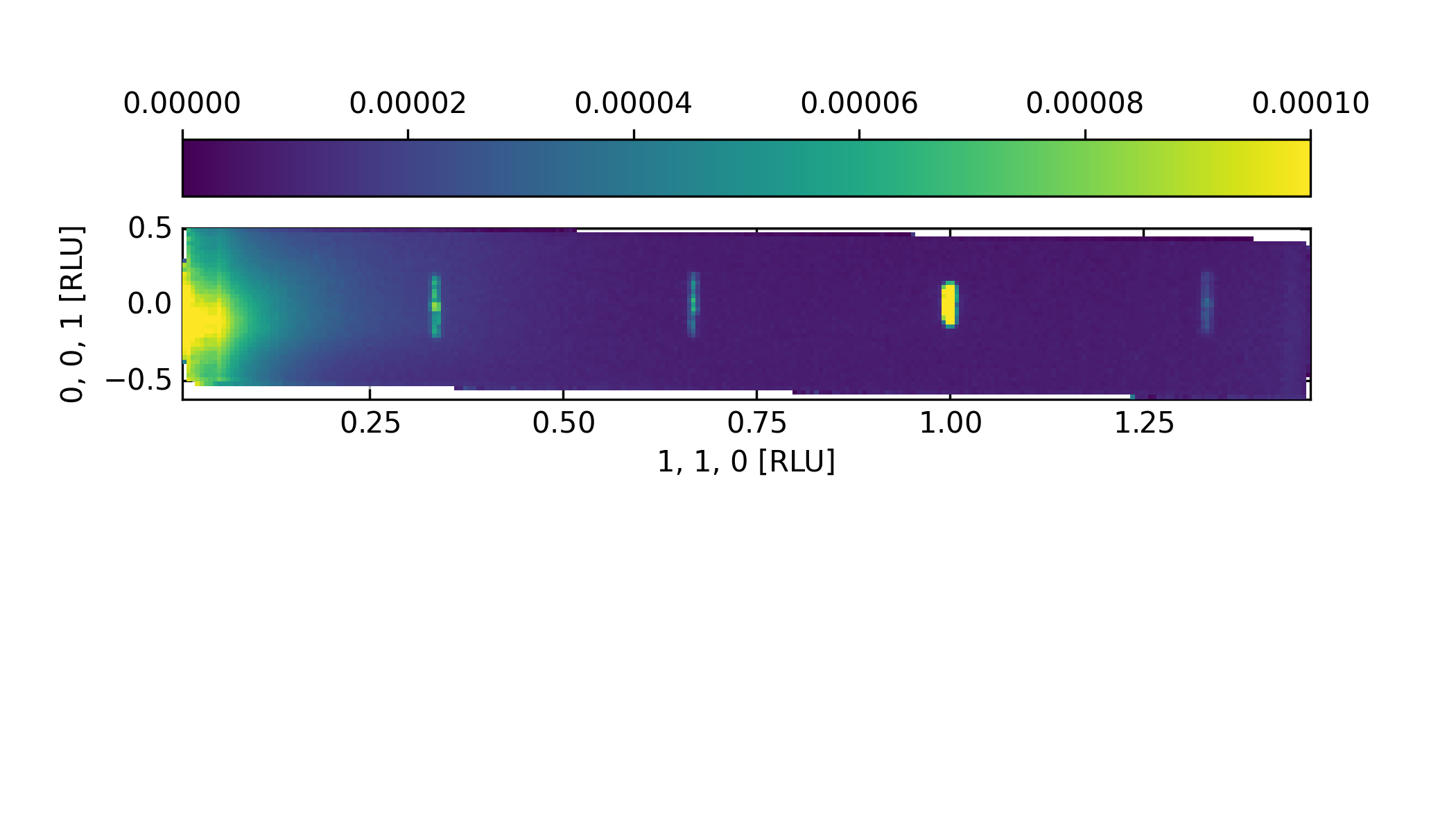}
    \caption{2D cuts performed with the \code{plotQplane}-function. Top: The HK0-scattering plane of a hexagonal sample. Bottom: A cut orthogonal to the measured scattering plane, corresponding to the HHK-plane. The \code{RLUAxes} ensures that the correct HKL-directions and angles are used for the figures. }
    \label{fig:2Dcut}
\end{figure}

A scattering plane is uniquely defined by three points in reciprocal space, $P_i$ for $i \in {1,2,3}$. From these points one can find the two main projection vectors within the scattering plane by i.e. $Q_1 = P_2-P_1$ and $Q_2 = P_3-P_1$. With these at hand, a rotation matrix transforming the data from the instrument coordinate system into the 2D plane system is generated, i.e. it transforms $Q_1$ to lie along the x-axis, and $\hat{Q}_3 = Q_1\times Q_2$ along the z-axis, and finally $\hat{Q}_2 = \hat{Q}_3\times Q_1$ to the y-axis. However, in addition to this matrix, and offset along $\hat{Q}_3$ has to be calculated when a 2D cut is requested which does not include the origin. From here, data is binned into the provided Q-grid for points inside $\pm 0.5$ width along $\hat{Q}_3$. Two examples, one with the plotted plane coinciding with the measured scattering plane and one orthogonal hereto, see Figure~\ref{fig:2Dcut}.

\subsubsection{1D Cutting}
\begin{figure}[ht!]
    \centering
    \includegraphics[width=0.6\linewidth]{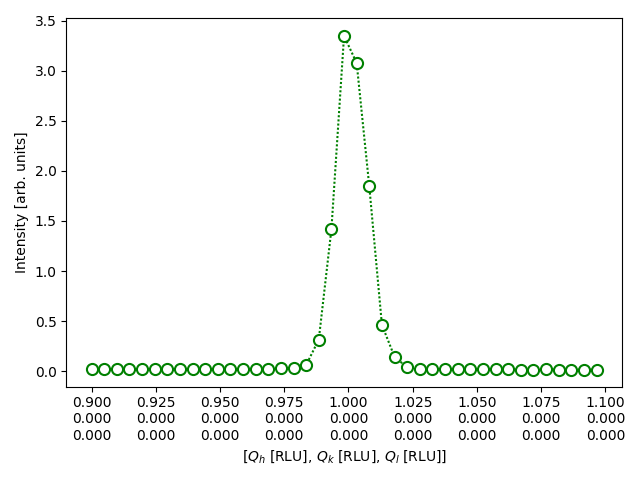}
    \caption{ Cut through a single crystal peak using the \code{plotCut1D} function. \code{generate1DAxis} is used to generate axes that display H, K and L at the same time.   }
    \label{fig:1Dcut}
\end{figure}

When performing 1D cuts through data, the straight forward method is to define a cylinder connecting the wanted Q points with a width defined from the cut width. For the number of data points in a standard DMC single crystal data file, this method is rather slow, and especially for dealing with the integrated intensities of single crystals, this function will be called up to several thousand times. In an effort to optimise the 1D cutting a bounding box is found from the provided start and end points and together with the cut width (expanded by 15 \% to ensure full coverage). Scattering angles and vertical detector position, i.e. sample rotation and scattering angle and vertical distance, are calculated for points along all edges of this bound box. This gives a quick masking in the natural coordinates of the data, i.e. sample rotation and position on the detector. An example of an extracted 1D cut is shown in Figure~\ref{fig:1Dcut}. Especially for short 1D cuts, this optimisation is effective. 

\subsection{Subtraction}
Subtraction of datasets is important to remove background from e.g. sample holders or identify diffuse scattering by subtracting a high-temperature dataset from a low-temperature dataset. Due to the RAM-optimization of not keeping all data file values in memory, in order to perform a subtraction between two datasets, a third needs to be created. The most utilized method is the \code{directSubtractDS}, which makes a direct subtraction of datasets, datafile by datafile, and requires the same exact scans to be performed, i.e. $A_3$ values of the corresponding scans have to be the same, and that the datasets are ordered in the same way. When monitor values differ, the highest monitor is chosen with the other data file intensity being rescaled accordingly.

\subsection{Single Crystal Validation}
As shown in the previous sections, the core functionality of single crystal data treatment is available within the \DMCPY{} software and allows the user to both perform peak integration as well as investigations of broad and diffuse signals. These features have been used in a couple of publications \cite{Andriushin2024,Littlehales2024}, . Additional showcasing is available within the data treatment illustrated in the online tutorials \cite{DMCPyReadTheDocs}, demonstrating the range of capabilities of \DMCPY{}.

\subsection{Graphical User Interface - \texorpdfstring{\code{DMCS.py}}{DMCS.py}}
\begin{figure}
    \centering
    \includegraphics[width=0.95\linewidth]{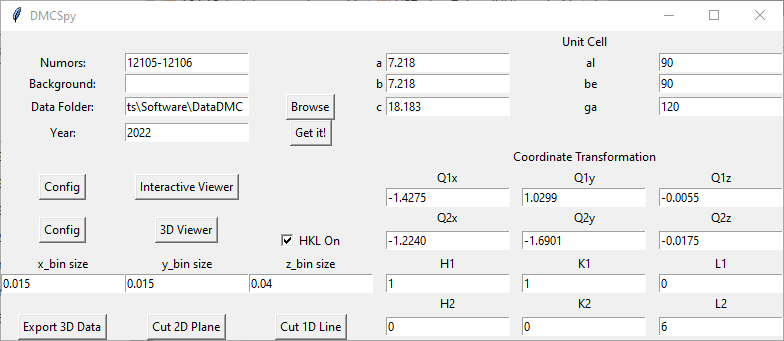}
    \caption{Graphical user-interface \code{DMCS.py} utilised on the instrument for a quick overview of data and to find the UB matrix. In this example, data from Cut2D HHL tutorial \cite{DMCPyReadTheDocs} is used with unit cell parameters and indexed peaks for alignment.}
    \label{fig:DMCSpy}
\end{figure}
To facilitate a quick and user-friendly interaction with the software during experiments on the instrument computer, a graphical user-interface has been created. It aids users in gaining a quick overview of data as well as to perform peak indexing by hand. Having two such indexed peaks allows for the creation of the UB matrix, which subsequently is used to generate cuts and plots along crystal symmetry directions. In addition, data can be exported. This interface is not intended to provide all of the needed functionality for a proper data analysis, but rather as an on-the-fly interface for data curation and initial data familiarisation. 

\FloatBarrier{}
\section{Conclusion}
Here, we have presented the \DMCPY{} code base used at the DMC diffractometer beam line at the Paul Scherrer Institute, Villigen, Switzerland. The two different modes of operation, powder and single crystal diffraction, have been introduced with an emphasis on the \DMCPY{} capabilities for single crystal data treatment, while the validation of the powder capabilities is shown through the publication utilising it. Lastly, the graphical user-interface \code{DMCS.py}, which serves as a quick tool during experiments, has also been introduced. Despite the tool being fully utilized in the user program at SINQ, a series of extensions and additional features could be implemented into it. These include support for goniometers/four-circle setup, integration with the instrument control software to easily choose regions of interest as well as possible further code optimizations in general. 

\section{Conflict of Interest}
We wish to confirm that there are no known conflicts of interest associated with this publication.

\section{Funding}

This research was founded by the Danish Agency for Research and Innovation through DanScatt grant 7055-00007B, the Research Council of Norway grant no. 325345 and the Norwegian Center for Neutron Research, NcNeutron, grant no. 245942.

\section*{Acknowledgements}
We are very thankful for the input and discussions during the development of the software with Lukas Keller, Oksana Zaharko and Ivan Usov. Further, we are grateful for the feedback in testing the software specifically from Amirreza Hemmatzade, Abraham Hernandez, Xavier Boraley, Manisha Islam, David Tam, and Tom Fennell.

\bibliography{bibliography}
\appendix

\newpage

\section*{Automatic Peak Search}\label{sec:PeakSearch}
To facilitate the finding and indexing of peaks used for a UB matrix, an automated alignment algorithm has been created. It utilized 10 steps to predict the correct scattering plane for a single crystal.
\begin{enumerate}
    \item Data is binned into equi-sized bins with a suitable size (dx,dy,dz)
    \item Potential peaks are identified by a simple thresholding
    \item A clustering is performed for points within 0.02 1/\AA and a centre of gravity is calculated
    \item Previous step is repeated for a user-defined distance allowing higher mosaicity crysaline samples
    \item Scattering plane normals are found by calculating cross products between found peaks
    \item The plane normals are clusted as in point 3 and most common plane normals is utilized as current plane normal
    \item All peaks are rotated such that the scattering plane normal is along $z$
    \item Peaks within the plane are projected along a list of most common reciprocal lattice scattering vectors, choosing the closest to an interger is chosen
    \item The identified peak is rotated to be correctly located within the scattering plan
    \item The rotation matrix to rotate the identified peak back to its original position is found and defines the inverse UB matrix
\end{enumerate}
The above method relies on the fact that when the cross product between Bragg peaks from within the same scattering plan is found, the length of the cross product is a multiple of the orthogonal direction. This is used to find the scattering plan within which most Bragg peaks lie.
Alternatively, one can manually look through the data, using e.g. the Viewer3D introduced in section \ref{sec:Viewer3D}, and perform the indexing of Bragg peaks. With this information, the rotation matrices as described above are found and the UB matrix created. It is here noted that no UB matrix optimisation is performed, i.e. lattice parameters are not refined in this process. 

\section*{Current executable software version}
\label{sec:Currentexecutablesoftwareversion}

Ancillary data table required for sub version of the executable software: (x.1, x.2 etc.) kindly replace examples in right column with the correct information about your executables, and leave the left column as it is.

\begin{table}[!ht]
\begin{tabular}{|l|p{4.3cm}|p{8.3cm}|}
\hline
\textbf{Nr.} & \textbf{(Executable) software metadata description} & \textbf{DMCpy} \\
\hline
S1 & Current software version & 1.0.1\\
\hline
S2 & Permanent link to executables of this version  & N/A   \\
\hline
S3 & Legal Software License & Mozilla Public License 2.0 (MPL-2.0)\\
\hline
S4 & Computing platforms/Operating Systems & Linux, OS X, Microsoft Windows, Unix-like \\
\hline
S5 & Installation requirements \& dependencies & Python  \\
\hline
S6 & If available, link to user manual - if formally published include a reference to the publication in the reference list & \url{https://dmcpy.readthedocs.io/} \\
\hline
S7 & Support email for questions & \url{mjolnirpackage@gmail.com}  \\
\hline
\end{tabular}
\caption{Software metadata (optional)}
\label{tab:SoftwaremetadataAppendig} 
\end{table}

\end{document}